# Density of states effects on emission and scattering of photons in plasmas

Sergey V. Gaponenko[1], Denis V. Novitsky[1], and Dmitry V. Guzatov[2]

[1] B. I. Stepanov Institute of Physics, National Academy of Sciences, Minsk 220072, Belarus
[2] Yanka Kupala State University of Grodno, Grodno 230023, Belarus

E-mail: s.gaponenko@ifanbel.bas-net.by

## Abstract

Plasma supports electromagnetic waves propagation for frequencies higher than plasma frequency but features dielectric permittivity less than 1. This property leads to photon density of states (DOS) lower than in vacuum and should result in subnatural spectral linewidths, sub-Planckian spectrum of thermal radiation, and sub-Rayleigh scattering as well as in lower inelastic photon scattering including Raman scattering. Lamb shift will be altered as well though the decisive contribution from high-energy modes in this case makes the photon DOS effect rather small since plasma DOS converges with the vacuum one in the limit of infinite frequencies. We emphasize the basic character of all these phenomena though absolute values of corrections in many real experiments may appear to be small as compared to other factors. We found that dissipative losses make possible DOS effects smaller though not vanishing and additionally bring about indefinite growth of DOS in the low-frequency limit.

## 1. Introduction

Photon density of states (DOS) in homogeneous space $D(k) = k^2/\pi^2$ with $k$ being the wave number of the mode in which photon is supposed to be spontaneously emitted or scattered defines probabilities of spontaneous transitions with photon emission, spectral density of thermal radiation, as well as probabilities of both elastic and inelastic photon scattering by quantum systems [1,2]. In vacuum its dependence on frequency $\omega$ when the dispersion law $\omega = ck$ holds ($c$ is the speed of electromagnetic radiation in vacuum) obeys the familiar relation,

$$D_0(\omega) = D(k)\frac{dk}{d\omega} = \frac{\omega^2}{\pi^2 c^3}. \qquad (1)$$

In this form photon DOS can be identified in the known expressions for spontaneous transitions probabilities (Einstein's relations), spectral density of thermal radiation (Planck's law), elastic scattering (Rayleigh formula) and inelastic scattering (Placzek formula) of photons. In other homogeneous media Eq. (1) changes because of the different dispersion law $\omega(k)$.

For media with local inhomogeneities of dielectric permittivity $\varepsilon(\omega,\mathbf{r})$ photon local DOS is to be introduced for every singularity which can be calculated using Green functions technique [3]. It is also possible to calculate spontaneous transitions probabilities using classical electrodynamics and the correspondence principle, i. e., the probability of spontaneous photon emission by a quantum system is believed to vary in accordance with inhibition or enhancement factor for electromagnetic power emitted by a classical oscillator at the same point $\mathbf{r}$ corresponding to $\varepsilon(\omega,\mathbf{r})$ local singularity or proximity of an antenna [4]. The very idea of spontaneous emission rate control by photon local DOS engineering dates back to the pioneer paper by E. M. Purcell in 1946 [5], and its extension to photon scattering has been suggested in [6]. Engineering of media with modified DOS (photonic crystals [7,8], metamaterials [9-12]) and structures with high local DOS (high-Q microcavities [13], optical antennas [14-16], and plasmonic structures [17-20]) constitutes an essential trend in modern optics and photonics [21,22]. However, the fundamental origin of the above phenomena makes their analogies possible in other ranges of electromagnetic waves beyond the optical one. Plasmas are known to feature variable dielectric function and its singularities in different frequency ranges of electromagnetic spectrum depending on charge concentration and its distribution in space. Recently we have shown that specific dispersion law inherent for electromagnetic waves in plasmas makes tunneling of electromagnetic radiation through a plasma layer to differ principally from that for dielectric structures[23]. In this work we highlight a number of possible basic physical effects resulting from photon DOS modification in plasmas.

## 2. Results and Discussion

*2.1. Ideal (lossles) plasma*

<u>Density of states in plasma</u>. The plasma dielectric function in the simple case obeys the known relation,

$$\varepsilon(\omega) = 1 - \frac{\omega_p^2}{\omega^2}, \ \omega_p^2 = \frac{Ne^2}{m\varepsilon_0}, \qquad (2)$$

where $\omega_p$ is referred to as plasma frequency, $N$ is electron density, $e$ is the elementary charge, $m$ is electron mass, and $\varepsilon_0$ is the dielectric constant. Eq. (2) holds when contribution from ions is neglected because of their high mass as compared to electrons, when collisions are absent and energy dissipation vanishes so that plasma response to electromagnetic radiation can be characterized by a real function. In many cases this simple relation can be used in a rather wide frequency range up to $\hbar\omega \ll mc^2$ [24], i. e., up to frequencies of the order of $10^{20}$ Hz. For $\omega > \omega_p$ the dielectric function is positive, $\varepsilon(\omega) > 0$, and plasma represents a dielectric supporting electromagnetic waves propagation with $\varepsilon < 1$ and the dispersion law,

$$\omega = (c^2 k^2 + \omega_p^2)^{1/2}, \qquad (3)$$

which ensures the condition $d\omega/dk < c$ to be met. Substitution of Eq. (3) into (1) gives the plasma density of states,

$$D(\omega) = \frac{\omega}{\pi^2 c^3}\sqrt{\omega^2 - \omega_p^2} =$$

$$= D_0(\omega)\frac{1}{\omega}\sqrt{\omega^2 - \omega_p^2} = D_0(\omega)\sqrt{\varepsilon(\omega)} \qquad (4)$$

This function is plotted in Fig. 1 along with the vacuum DOS (1). The DOS function (4) differs from its vacuum counterpart (1) by the factor which tends to 1 for ω→∞ and tends to zero for ω→$\omega_p$.

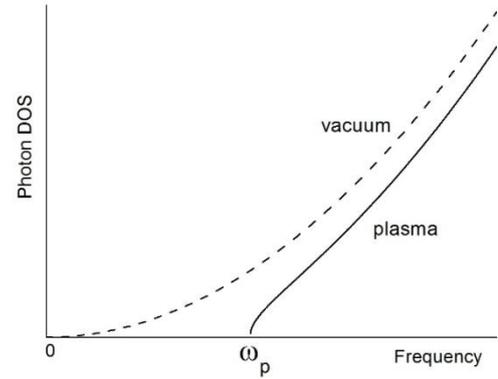

Fig. 1. Density of photon states versus frequency in vacuum and in plasma.

Depletion of photon DOS in plasmas versus vacuum will result in modification of every process which is defined by photon DOS. These are [2] spontaneous decay rate, Planck law for thermal radiation, Rayleigh and Raman photon scattering, and Lamb shift. All these processes bear the photon DOS factor in equations describing their rate/intensity/probability/ Therefore plasma effects on these processes have fundamental origin, however for these to become apparent and detectable it is necessary that (i) concentration dependent plasma frequency should not be negligibly small as compared to the frequency of interest for a physical phenomenon in question, and (ii) the corrections from the specific plasma property 0< $\varepsilon(\omega)$ <1 should be high enough to be detectable in presence of other affecting factors in every experimental situation. Let us consider a few selected cases.

<u>Subnatural spectral linewidths in plasmas</u>. In proportion with photon DOS decrease in plasmas, spontaneous transitions rate will be inhibited, i. e. slowing down of spontaneous decay should occur and the corresponding intrinsic linewidth will change accordingly. That means atoms and ions should exhibit subnatural spectral linewidths. Spectral width narrowing with respect to the natural linewidth looks rather difficult for experimental observation because of the Doppler broadening of emission lines resulting from random motion of atoms and ions in plasmas. Doppler broadening typically dominates over natural widths and makes subnatural lines narrowing nondetectable. Most probably, ultracold plasmas where ionization occurs owing to direct laser excitation rather than heating enable detectability of subnatural linewidth. The typical electron concentration in ultracold plasmas are of the order of $10^{10}$ cm$^{-3}$, the temperature being as low as 1-10 K [25]. This concentration corresponds to $f_p = \omega_p/2\pi = 0.9$ GHz, i.e., to the microwave range. E. g., the 10% narrowing of atomic linewidth versus the natural one will hold for frequency $f \approx 2$ GHz. One more interesting phenomenon may appear upon $\omega \to \omega_p$. In this case strong variation in photon DOS may occur across emission spectrum, i. e., inside the spectral line under consideration. In this case the simple consideration in terms of transition probability may appear to be not relevant and non-Markovian relaxation of an excited quantum system may become the case [26]. Furthermore, photon DOS tending to zero for $\omega \to \omega_p$ may cause an overall inhibition of spontaneous decay similar to the case of photonic crystals [7,27].

Modified Planck law for plasmas. Let us consider modification of the fundamental Planck law for thermal radiation. The spectral energy density for thermal radiation $u(\omega)$ is treated as the product of the photon DOS, the distribution function describing occupation of available states by photons, and the photon energy to give the known relation

$$u(\omega) = \hbar\omega \frac{\omega^2}{\pi^2 c^3}(exp\frac{\hbar\omega}{k_B T} - 1)^{-1} \qquad (5)$$

with $k_B$ being the Boltzman constant. This relation is valid for vacuum and for every medium different from vacuum the explicit factor describing photon DOS should be used. Deviation of thermal radiation from the Planck law is the subject of extensive discussion with respect to various media and nanostructures for the purpose of thermal emittivity control and optimization [28-34]. With photon DOS given by Eq. (4) we arrive at the modified Planck formula for plasma,

$$u_{plasma}(\omega) = \hbar\omega \frac{\omega}{\pi^2 c^3}\sqrt{\omega^2 - \omega_p^2}(exp\frac{\hbar\omega}{k_B T} - 1)^{-1} \quad . \quad (6)$$

Eq. (6) gives pronounced inhibition of thermal enmission for lower frequencies. For further discussion let us move from the energy density versus frequency to energy density versus wavelength in vacuum $\lambda_0$ as is usually used in experiments. For vacuum, using the relations

$$u(\lambda_0) = u(\omega)\frac{d\omega}{d\lambda_0}, \quad \omega = 2\pi c/\lambda_0 \qquad (7)$$

one arrives at the textbook equation

$$u(\lambda_0) = \frac{8\pi hc}{\lambda_0^5}(exp\frac{hc}{\lambda_0 k_B T} - 1)^{-1}. \qquad (8)$$

The same approach using Planck formula for plasma (6) gives the modified $u_{plasma}(\lambda_0)$ expression,

$$u(\lambda_0) = \frac{8\pi hc}{\lambda_0^5}\sqrt{1 - \frac{\lambda_0^2}{\lambda_p^2}}(exp\frac{hc}{\lambda_0 k_B T} - 1)^{-1}, \lambda_p = 2\pi c/\omega_p \quad (9)$$

with $\lambda_p$ being the radiation wavelength in vacuum relevant to plasma frequency value $\omega_p$. Fig. 2 presents both vacuum and plasma Planck laws. For convenience the emitted power per unit square per solid angle $I(\lambda_0) = u(\lambda_0)c/4\pi$ is plotted to get results in Watt per square meter instead of Joule per cubic meter. One can see that plasma inhibits thermal emission at the long-wave wing of the spectrum disabling emittivity completely for $\lambda_0 > \lambda_p$.

Notably, using vacuum wavelength is principally important here since we suppose that plasma properties are examined by an observer located in the vacuum-like ambient medium but not in plasma. If the in situ experiment is thought, i. e., a detector is located directly inside plasma, then Planck law will further change in accordance with

$$u_{plasma}(\lambda) = u_{plasma}(\lambda_0)\frac{d\lambda_0}{d\lambda}, \lambda = \lambda_0(1 - \lambda_0^2/\lambda_p^2)^{-1/2}. \qquad (10)$$

Wavelength $\lambda$ in plasma can be expressed as in (10) or as

$$\lambda = \lambda_0(1 - \omega_p^2/\omega^2)^{-1/2} = \lambda_0/\sqrt{\varepsilon(\omega)} \quad . \quad (11)$$

It differs from $\lambda_0$ because of the specific dispersion law (3) which should be used to derive $d\lambda_0/d\lambda$ in Eq. (10). Finally, we arrive at the relation for thermal energy density u in plasma versus radiation wavelength measured in plasma $\lambda$ which reads,

$$u_{plasma}(\lambda) = \frac{8\pi hc}{\lambda^5}\sqrt{1 + \frac{\lambda^2}{\lambda_p^2}}(exp\frac{hc\sqrt{1+\lambda^2/\lambda_p^2}}{\lambda\ k_B T} - 1)^{-1},$$

$$\lambda_p = 2\pi c/\omega_p \qquad (12)$$

Notably this in situ presentation of the modified Planck law for thermal radiation in plasma merges with the relevant relation for vacuum (7) for large values of $\lambda_p$ (i. e., for low plasma concentrations) but does not feature any cut-off in the spectrum. The cut-off is not present here since when vacuum wavelength $\lambda_0$ approaches plasma wavelength $\lambda_p$, the radiation wavelength $\lambda$ in plasma defined by (10) and (11) infinitely grows up. Thus, in this Gedankenexperiment in situ an observer will see the continuous spectral dependence of $u_{plasma}(\lambda)$.

Let us estimate possible detectability of sub-Planckian radiation from the Sun (the insert in Fig. 2(b)). Taking plasma concentration to be about $10^{17}$ cm$^{-3}$ at the surface one has $\lambda_p = 107$ μm which gives 1%, 10%, and 50% drop in intensity versus the Planck law for wavelengths of 15, 46, and 86 μm, respectively.

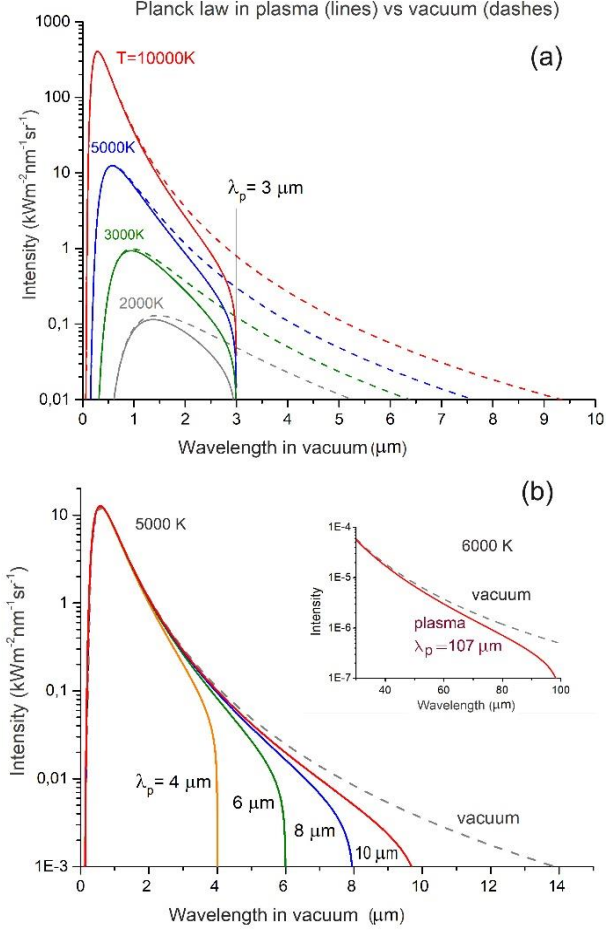

Fig. 2. Planck law for thermal radiation in plasma (lines) and in vacuum (dashes) versus radiation wavelength in vacuum λ0 for (a) various temperatures and (b) various plasma wavelength λp. Insert in (b) shows the portion of Planck law for λp = 107 μm which is supposed to be plausible for solar radiation.

Yet another case of Planck law violation can be traced in fusion experiment with plasma in tokamak reactors where the target plasma concentration reaches 1014 cm-3 with temperature exceeding 106 K [35]. In this case strong deviations from Planck law should be pronounced in the microwave or even terahertz range for radiation wavelength in vacuum of the order of 3 mm.

Much higher plasma concentrations are feasible in electron-hole semiconductor plasma at hard laser impact by ultrashort laser pulses. Taking concentration to be about $10^{21}$ cm$^{-3}$ [36] one has $f_p = \omega_p/2\pi = 2.8 \cdot 10^{14}$ Hz, and 1% deviation from the Planck law at $2 \cdot 10^{15}$ Hz, and 10% drop at $6.5 \cdot 10^{14}$ Hz corresponding to the UV range: 150 and 460 nm, respectively. Notably, thermal radiation will not be emitted at all below plasma frequency. For the above conditions, the cut-off wavelength will be $\lambda_p = c/f_p = 1.1$ μm. In these examples, wavelengths are given for vacuum and to be scaled in plasma in proportion to $\varepsilon^{-1/2}$. This condensed matter example is relevant to ultrashort laser pulses applied for semiconductor rapid heating and annealing.

It should be noted that the lower values of dielectric function in plasma cannot alone enable all effects highlighted here but the specific dispersion law inherent in plasma and resulting in certain *dk/dω* function is also essential. There exist media with low permittivity with the reverse effect, namely spontaneous decay may experience enhancement for lower permittivity as is the case in metamaterials and certain waveguides with *n* < 1 [26,37,38].

*Lamb shift in plasmas.* Modified photon density of states will definitely alter the value of Lamb shift which represents the distinctive manifestation of electromagnetic vacuum fluctuations. Change of photon DOS in plasma with respect to vacuum will result in lower energy of zero oscillations and should result in the lower Lamb shift accordingly. Modification of Lamb shift has been in fact considered for certain structured dielectrics [39-41] raising a discussion whether the effect can really be observable. The peculiarity of the Lamb shift physics is the decisive role of high-energy modes up to enormous numbers close to $\hbar\omega_{max} = mc^2$ [42]. Neither dielectric suggests measurable change in photon DOS in this high-frequency range. Our calculations made using formalism elaborated in [41,43] show that in fact, in spite that plasma does feature photon DOS modification for all photon energies, its asymptotic convergence with vacuum DOS for infinite frequency growth results in the photon DOS correction to Lamb shift no more than 1% at plasma densities up to $10^{22}$cm$^{-3}$ (see Fig. 3), the huge density which is feasible only in condensed matter and possibly in certain astrophysical objects.

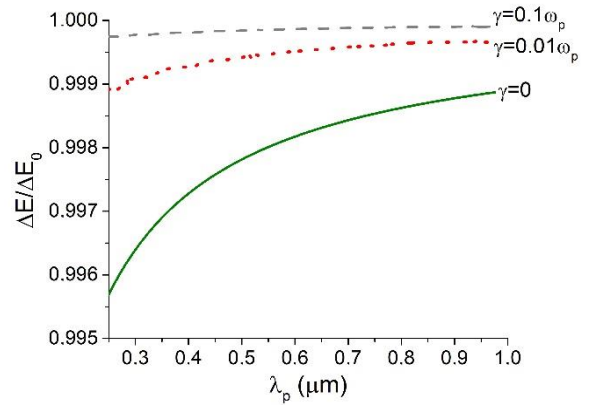

Fig. 3. Calculated Lamb shift in plasma ΔE normalized to its vacuum value ΔE0 as the function of the plasma wavelength λp defined by Eq. [9]. The γ parameter describes losses.

In Fig. 3 we show the results of calculations for an ideal plasma without losses (denoted as γ = 0). Finite losses make Lamb shift modification in plasma even lower. Therefore, we summarize that Lamb shift modification in plasmas with respect to vacuum though being the fundamental physical phenomenon gains extremely low values and presents a challenge for experimental detection.

## 2.2. Plasma with dissipation losses

It should also be realized that for all the effects highlighted here to occur there is no necessity the all conditions used to derive Eq. (2) to be met. These conditions make consideration simple, apparent and straightforward. It is clear however, that DOS effects in plasma will occur when dissipation cannot be absolutely neglected. In this context, possible modification of electron-positron annihilation rate in dense astrophysical plasmas can become the subject of analysis [44] as well as possible modification of nuclear processes provided these occur with photon emission (in the gamma-range) especially in quark-gluon plasma.

The causality principle is known to result in certain constraints on the total control of spontaneous emission rate at a given point of space. Relative changes in density of states when integrated over frequency range from zero to infinity gives finally zero as has been proved for non-absorbing dielectrics by Barnett and Loudon [45] and then has been extended to absorbing dielectrics by Scheel [46]. One can readily observe that Eq. (4) does not meet this rule since it predicts only inhibition of spontaneous decay rate and this is because Eq. (2) does not conform with Kramers—Kronig relations implying that dispersion can exist only along with absorption. Therefore, to be more realistic, one should take absorption into account. In the general case of an absorbing medium, the DOS formula (4) should be changed to [47]

$$D(\omega) = D_0(\omega)n(\omega) = D_0(\omega)Re\sqrt{\varepsilon(\omega)}, \qquad (13)$$

where $n(\omega)$ is the refractive index calculated from the complex-valued permittivity $\varepsilon(\omega)$. As an example, let us consider the straightforward generalization of the Drude model with

$$\varepsilon(\omega) = 1 - \omega_p^2/\omega(\omega + i\gamma), \qquad (13)$$

where $\gamma$ is the dissipation parameter. The real and imaginary parts of this permittivity for different $\gamma$ are shown in Fig. 4(a). The resulting refractive index and, hence, modification of DOS is demonstrated in Fig. 4(b). One can see that for $\gamma > 0$, the DOS does not vanish below the plasma frequency but grows indefinitely at ω→0. This fact is essential for the results to be consistent with the Barnett—Loudon—Scheel sum rule [45,46]

$$\int_0^\infty \left[\frac{D(\omega)}{D_0(\omega)} - 1\right] d\left(\frac{\omega}{\omega_p}\right) =$$

$$= \int_0^\infty [n(\omega) - 1]d(\omega/\omega_p) = 0, \qquad (14)$$

governing the overall compensation of DOS decrease at certain frequencies with its increase in another frequency range. For $\gamma = 0$, the integral on the left-hand side of Eq. (14) gives the nonzero value $-\pi/2$. Nevertheless, since we are interested mostly in the frequency range above $\omega_p$, the low-DOS effects discussed above for the lossless media should remain essentially the same for $\gamma > 0$ as well.

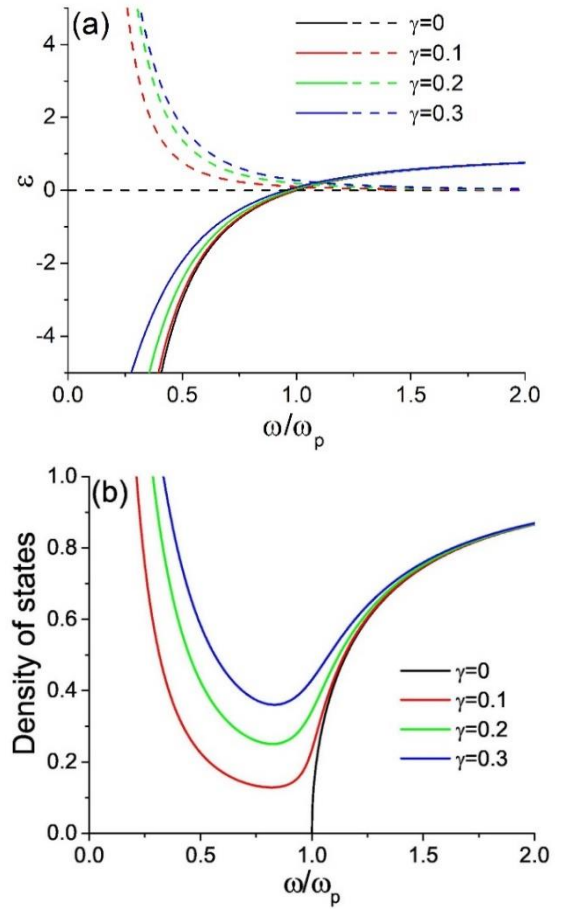

Fig. 4: Frequency dependencies of (a) permittivity (solid lines for real parts, dashed lines for imaginary parts) and (b) DOS.

Hitherto, plasma has been treated as an infinite continuous medium since our goal was to reveal the principal possible effects originating from unusual dielectric permittivity of plasmas versus typical dielectric media. In real experiments, presence of the border (e.g. plasma/vacuum or plasma/dielectric) or even confinement (a plasma layer between two dielectric or metal walls) phenomena may affect considerably both photon emission/scattering and radiation extraction processes.

## 2.3. Interface effects

Consider the effect of a plasma/dielectric border which can modify radiation propagation and extraction into a dielectric medium. If a semi-infinite plasma half-space borders with a dielectric half-space, say a vacuum, then using common relations [48] for transmission coefficient of a finite layer whose permittivity obeys Eq. (13) and which is confined between two half-spaces with permittivity obeying Eq. (13) and $\varepsilon = 1$, one can arrive at the relation for the layer $h$ thickness relevant to $1/e$ intensity decay versus normalized to the vacuum wavelength $\lambda_0$,

$$\frac{h}{\lambda_0} = \frac{1}{4\pi Im\sqrt{\varepsilon}}\left[1 + ln\frac{4|\sqrt{\varepsilon}|}{|\sqrt{\varepsilon}+1|^2}\right].\quad(15)$$

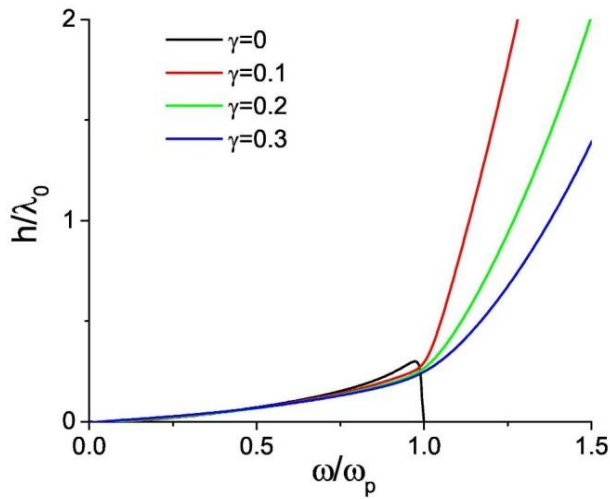

FFig. 5. Calculated $h$ depth for $1/e$ intensity of electromagnetic radiation decay for plasma/vacuum interface normalized to vacuum wavelength $\lambda_0$.

The results are plotted in Fig. 4. One can see that in the ideal case of lossless plasma ($\gamma = 0$) the characteristic $h$ thickness rapidly falls to 0 for frequency approaching the $\omega_p$ value (i.e. when permittivity approaches zero). That means radiation extraction into adjacent dielectric medium will be strongly inhibited below the plasma frequency. Note, this is well known situation in optics for metal films where the characteristic $1/e$ depth is referred to as the skin layer. In the lossless case plasma readily supports transmission of electromagnetic radiation and there is no need to introduce $h$ value at all, but for the case $\varepsilon = 0$ radiation extraction from plasma into dielectric becomes questionable and appears to be beyond the subject of the present work. In case of finite dissipation described by the $\gamma = 0.1…0.3$ radiation extraction outside plasma looks feasible at all frequencies, the characteristic decay length being equal to a few $\lambda_0$ lengths for frequencies higher than plasma frequency. Generally speaking, observation of the predicted phenomena of photon density of states in plasma/dielectric bordering media may become non-trivial when simultaneously the two limits $\omega \to \omega_p$ and $\gamma \to 0$ hold.

Even more non-trivial situation may be foreseen when a finite plasma layer is confined by reflecting surfaces serving as mirrors. Then cavity quantum electrodynamics will be the relevant conceptual platform and unusual plasma properties are to be accounted correctly for modes consideration in a cavity [13]. This subfield has not received much attention to date [49] but may become the subject of more active research in the nearest future.

Finally, if a 3D-confinement of plasma is performed, i. e., a portion of plasma is somehow isolated in a dielectric space, then in the case of the size restrictions comparable with the wavelength one can deal with the conceptions of antenna known both in radiophysics and nowadays in optics with the realm of light—matter interaction phenomena in action, the field is being the subject of plasmonics (see, e. g., [20, 21] and references therein) in which the *local* photon density of states becomes the principal figure of merit in describing space inhomogeneities with respect to radiation emission and scattering.

## 3. Conclusions

Plasmas are suggested to feature lower photon density of states (DOS) as compared to vacuum because of specific dielectric function $\varepsilon(\omega)$ and therefore should exhibit modified spontaneous emission rates, thermal radiation spectral density, elastic and inelastic photon scattering as well as Lamb shift. The plasma DOS versus vacuum DOS factor equals $[\varepsilon(\omega)]^{1/2} < 1$ in an ideal case of negligible absorption loss which should result in slowing down of spontaneous radiative decay, subnatural spectral linewidths, sub-Planckian thermal radiation and inhibited elastic and inelastic photon scattering for frequency higher than plasma frequency as well as in lower Lamb shift. Account for absorption with $Re\{[\varepsilon(\omega)]^{1/2}\}$ as plasma DOS versus vacuum DOS factor gives not only DOS below vacuum DOS for frequencies higher than plasma frequency but also strong enhancement of DOS and all related processes in the limit of low frequencies.

## Acknowledgements

Helpful discussions with N. V. Tarasenko, G. V. Vereschagin, M. B. Shundalov, and A. S. Garkun are acknowledged. The work has been supported by the State Program for Scientific Research "Photonics and Electronics for Innovations" (Task 1.5).

**References**

[1]  Novotny L and Hecht B, *Principles of Nano-Optics* (Cambridge University Press, Cambridge) 2012.
[2]  Gaponenko S V, *Introduction to Nanophotonics* (Cambridge University Press, Cambridge) 2010.
[3]  D'aguanno G, Mattiucci N, Centini M, Scalora M, and Bloemer M J, 2004. *Physical Review E*, **69,** 057601.
[4]  Klimov V V and Ducloy M, 2004 *Phys. Rev. A* **69,** 013812.
[5]  Purcell E M, 1946, *Phys. Rev.* **69**, 681.
[6]  Gaponenko S V 2002 *Phys. Rev. B* **65**, 140303.
[7]  Bykov V. P., *Radiation of Atoms in a Resonant Environment* (World Scientific, Singapore) 1994.
[8]  Busch K, Von Freymann G, Linden S., Mingaleev S F, Tkeshelashvili L. and Wegener M 2007 *Physics Reports*, **444**, 101.
[9]  Cai W and Shalaev V M *Optical Metamaterials* (Springer, New York) 2010.
[10] Zheludev N 2016 *J. Optics* **18** , 1.
[11] Poddubny A., Iorsh I., Belov P. and Kivshar Y. 2013 *Nature Photonics*, **7**, 948.
[12] Sarychev A.K., Shalaev V.M. *Electrodynamics of Metamaterials.* (World Scientific, London, Singapore) 2007.
[13] Walther H., Varcoe B.T., Englert B.G. and Becker T., 2006. *Rep. Progr. Phys.* **69** 1325.
[14] Biagioni P, Huang J S, and Hecht B 2012 *Rep. Progr. Phys.* **75** 024402.
[15] Krasnok A E et al. 2011 *JETP Lett.* **94** 593.
[16] Bondarev I V, Slepyan G Y and Maksimenko S A, 2002 *Phys. Rev. Lett.* **89** 115504.
[17] Novotny L and Van Hulst N 2011 *Nature Photonics* **5** 83.
[18] Klimov V *Nanoplasmonics* (CRC Press) 2014.
[19] Vinogradov A P, Andrianov E S, Pukhov A A, Dorofeenko A V. and Lisyansky A A 2012 *Physics-Uspekhi* **55**, 1046.
[20] Gaponenko S V and Guzatov DV 2020 *Proc. IEEE* **108** 704.
[21] Barnes W L, Horsley S A and Vos W L 2020 *J.Optics*, **22** 073501.
[22] Gaponenko S V 2014 *J. Nanophotonics* **8** 087599.
[23] Gaponenko S V, Novitsky D V 2022 *Phys. Rev. A* **106** 023502.
[24] Ginzburg V. L. *The Propagation of Eelectromagnetic Waves in Plasmas* (*Int. Ser. Monographs in Electromagnetic Waves* (1970).
[25] Killian T C, Pattard T, Pohl T and Rost J M 2007 *Phys.Rep.* **449**, 77.
[26] Li Z.Y, Lin L L and Zhang Z Q 2000 *Phys. Rev.Lett.* **84** 4341.
[27] Mogilevtsev D, Moreira F, Cavalcanti S B, Kilin S. (2006) *Laser Phys. Lett*, **3** 327.
[28] Lin S.-Y., Fleming J. G., Chow E., Bur J., Choi K. K. and Goldberg A. 2000 Phys.Rev. B, 62 R2243.
[29] Maruyama S., Kashiwa T., Yugami H. and Esashi M. 2001 Appl. Phys. Lett. 79 1393.
[30] Guo Y., Cortes,C. L., Molesky S., Jacob, Z. 2012. Applied Physics Letters, 101, 131106.
[31] Nefedov I. S., Melnikov L. A. 2014 Applied Physics Letters, 105, 161902.
[32] Yu, Z., Sergeant, N. P., Skauli, T., Zhang, G., Wang, H., & Fan, S. 2013 Nature communications, 4, 1.
[33] Sokolsky, A. A., Gorlach, M. A. 2014 Physical Review A, 89, 013847.
[34] Baranov, D. G., Xiao, Y., Nechepurenko, I. A., Krasnok, A., Alù, A., Kats, M. A. 2019 Nature materials, 18, 920.
[35] Cesario, R., Amicucci, L., Cardinali, A., Castaldo, C., Marinucci, M., Panaccione, L., Santini, F., Tudisco, O., Apicella, M.L., Calabro, G. and Cianfarani, C., 2010. Nature Communications, 1, 55.
[36] Ashitkov S.I., Ovchinnikov A.V. and Agranat, M.B., 2004 *J. Ex.Theor. Phys. Lett.* **79** 529.
[37] So J K, Yuan G H, Soci C and Zheludev N I 2020. *Appl. Phys. Lett.* **117** 181104.
[38] Chebykin A V, Orlov A A, Shalin A S, Poddubny A N and Belov P A, 2015 *Phys. Rev. B* **91**, 205126.
[39] Zhu S Y, Yang Y, Chen H., Zheng H and Zubairy M S, 2000 *Phys. Rev. Lett.* **84** 2136.
[40] Wang X H, Kivshar Y S, Gu B Y 2004 *Phys. Rev. Lett.* **93** 073901.
[41] Li Z Y, Xia Y 2001 *Phys. Rev. B* **63** 121305(R).
[42] Bethe H A 1947 Phys. Rev. **72** 339.
[43] Matloob R 2000 *Phys. Rev. A* **61** 062103.
[44] Vereshchagin G V and Aksenov A G *Relativistic Kinetic Theory: With Applications in Astrophysics and Cosmology* (Cambridge University Press, Cambridge) 2017.
[45] Barnett S M and Loudon R 1996 *Phys. Rev. Lett.* **77** 2444.
[46] Scheel S 2008 *Phys. Rev.A* **78** 013841.
[47] Barnett S M, Huttner B and Loudon R 1992 *Phys. Rev. Lett.* **68** 3698.
[48] Born M and Wolf E 2013. *Principles of optics: electromagnetic theory of propagation, interference and diffraction of light*. Elsevier. (Section 1.6).
[49] Rokaj V, Ruggenthaler M, Eich F G, Rubio A 2022 *Phys. Rev. Res.* **4** 013012.